\documentclass[12pt]{article}
\pdfoutput=1
\usepackage{geometry,enumerate,amsmath,amssymb}
\usepackage{fullpage}
\usepackage{graphicx}
\usepackage{bm}
\numberwithin{equation}{section}
\newcommand{\be}{\begin{equation}}
\newcommand{\ee}{\end{equation}}
\newcommand{\bea}{\begin{eqnarray}}
\newcommand{\eea}{\end{eqnarray}}
\renewcommand{\epsilon}{\varepsilon}

\newcommand{\bphi}{\mbox{\boldmath $\phi$}}

\begin{document}
\title{Q-lump scattering}
\author{
  Paul Sutcliffe\\[10pt]
 {\em \normalsize Department of Mathematical Sciences,}\\
 {\em \normalsize Durham University, Durham DH1 3LE, United Kingdom.}\\ 
{\normalsize Email:  p.m.sutcliffe@durham.ac.uk}
}
\date{April 2023}

\maketitle
\begin{abstract}
  Q-lumps are spinning planar topological solitons with stationary solutions that satisfy first-order Bogomolny equations. Q-lump scattering has previously been studied only in the charge two sector, by approximating time evolution by motion in the moduli space of stationary solutions. In this paper, higher charge scattering is studied via motion on families of 4-dimensional submanifolds of moduli space, obtained by imposing cyclic symmetries. The results are shown to be in good agreement with field theory simulations, which are then applied to study more complicated Q-lump scattering processes, including examples where the moduli space approximation is not applicable. A variety of exotic scattering events are presented.  
  \end{abstract}

\

\newpage
\section{Introduction}\quad
Q-lumps are spinning topological soliton solutions of the $O(3)$ $\sigma$-model with an easy axis anisotropy in 2-dimensional space \cite{Le}. This theory has no static soliton solutions, but there are stationary Q-lump solutions in which an internal phase rotates at a constant angular frequency, given by the mass parameter of the theory. Q-lumps can have any size and they satisfy first-order Bogomolny equations that are solved explicitly in terms of based rational maps. For each positive integer $N$, this gives a $4N$-dimensional moduli space of stationary Q-lumps with topological charge $N$. This reflects the fact that there are no forces between stationary Q-lumps, with the parameters of moduli space being interpreted as a position, size and internal phase for each Q-lump, in the asymptotic region in which they are all well-separated. However, for $N=1$ all Q-lumps have infinite energy, because the field decays too slowly for the energy integral to converge. For $N>1$ the dimension of the moduli space of finite energy stationary Q-lumps is reduced to $4N-2$, due to a constraint that may be viewed in the asymptotic region as imposing conditions that relate some of the parameters of the constituent Q-lumps to each other.

Q-lump scattering has previously been studied only in the charge two sector \cite{Le} by applying the moduli space approximation \cite{Ma}, where dynamics is restricted to the moduli space of finite energy stationary solutions, equipped with a metric and a potential induced from the field theory Lagrangian. This allowed the investigation of the scattering of a pair of unit charge Q-lumps, but the Q-lumps are restricted to have equal sizes and are phase locked with an internal relative phase of $\pi$. Nonetheless, a rich structure was found for Q-lump scattering in the charge two sector, with features in common with both topological solitons in $\sigma$-models \cite{book} and Q-balls \cite{Co}, but with aspects not found in either of these soliton systems.
In the present paper, higher charge scattering is investigated using the moduli space approximation on families of 4-dimensional submanifolds of finite energy stationary Q-lumps, obtained by imposing cyclic symmetries. This allows the first study of Q-lump scattering with relative phases that are not locked at $\pi$, and also includes examples where Q-lumps have different sizes.

Field theory simulations have never been performed for Q-lump scattering. This situation is remedied here, where field theory computations are shown to be in good agreement with the results of moduli space dynamics. Field theory simulations are then applied to higher charge scattering, where the moduli space approach would be cumbersome due to the reasonably large dimensions of the moduli spaces involved. In particular, the scattering of a pair of charge two Q-lumps is found to yield exotic dynamics in the charge four sector. Finally, field theory simulations are applied to the scattering of Q-lumps where the moduli space approximation is not applicable, because the fields have infinite energy. A pair of unit charge Q-lumps with different sizes, or a relative phase not locked at $\pi$, being prototypical examples. It is found that the dynamics is similar to the finite energy case, indicating that considerations of finite energy are not particularly relevant in the study of local Q-lump dynamics.

\section{Q-lumps}\quad
The theory of interest for Q-lumps is the relativistic $O(3)$ $\sigma$-model in (2+1)-dimensions, modified by the addition of a symmetry breaking mass term \cite{Le}. The field
$\bphi=(\phi_1,\phi_2,\phi_3)$ is a 3-component unit vector with the Lagrangian density 
\be
   {\cal L}=\frac{1}{2}\partial_\mu\bphi\cdot\partial^\mu\bphi
   -\frac{m^2}{2}(\phi_1^2+\phi_2^2),
   \label{lag}
   \ee
   where the index $\mu\in\{0,1,2\}$ runs over the time and space coordinates. The mass term in (\ref{lag}) is also familiar as an easy axis anisotropy term from the continuum description of a ferromagnet, although the dynamics in that case is different, being first-order in time rather than the second-order relativistic system considered here. The positive constant $m$, giving the mass of the $\phi_1$ and $\phi_2$ fields, is taken to be $m=\frac{1}{20},$ but this value is not particularly significant as different values of $m$ can be related by a rescaling of the spacetime coordinates.

   The required boundary condition is $\bphi\to(0,0,1)$ as $x^2+y^2\to\infty$, thereby providing a compactification of space from $\mathbb{R}^2$ to $S^2$ by the addition of the point at infinity. At any given time, $\bphi$ is therefore a map between two-spheres and has an associated integer-valued topological charge due to the homotopy group formula $\pi_2(S^2)=\mathbb{Z}.$ This topological charge, $N$, can be calculated as the degree of the mapping and is given by the integral
   \be
   N=-\frac{1}{4\pi}
  \int\bphi\cdot(\partial_x \bphi\times\partial_y \bphi)\,dxdy.
\label{charge}
\ee

It is helpful to introduce the $\mathbb{CP}^1$ formulation of the model by using stereographic projection to define the Riemann sphere coordinate
   $W=(\phi_1+i\phi_2)/(1+\phi_3)$. In this formulation the Lagrangian density (\ref{lag}) becomes 
\be
   {\cal L}
   =\frac{2}{(1+|W|^2)^2}(\partial_\mu W\partial^\mu \overline W-m^2|W|^2),
   \label{cp1lag}
   \ee
   with a variation that yields the nonlinear field equation
   \be
   (1+|W|^2)\partial_\mu\partial^\mu W-2\overline W\partial_\mu W\partial^\mu W
   +m^2W(1-|W|^2)=0.
   \label{el}
   \ee
It is easy to check that stationary solutions of (\ref{el}) can be obtained by solving the first-order Bogomolny equations
   \be
   \partial_y W=\pm i\partial_x W, \quad \mbox{ and } \quad \partial_t W=\pm imW,
   \label{bog}
   \ee
   where the signs in the above can be chosen independently, but both are taken to be positive in the following.

   In the topological charge $N$ sector there is a $4N$-dimensional moduli space ${\cal M}_N$ of stationary solutions of the Bogomolny equations (\ref{bog}), called Q-lumps \cite{Le}, given by
   \be
   W=\frac{\alpha_{N-1}z^{N-1} +\cdots +\alpha_1 z+\alpha_0}
   {z^N+\beta_{N-1}z^{N-1} +\cdots +\beta_1 z+\beta_0}e^{imt},
   \label{ratmap}
   \ee
   where $z=x+iy.$ The complex constants $\alpha_i,\beta_i,$ for $i=0,..,N-1,$
   are coordinates on ${\cal M}_N$, and are subject only to the constraint that the numerator and the denominator in the based rational map that appears in (\ref{ratmap}) have no common roots. The rational map is based because the degree of the numerator is less than the degree of the denominator, due to the boundary condition that $\bphi\to(0,0,1)$ as $x^2+y^2\to\infty,$ which requires that $W\to 0$ as $|z|\to\infty.$

   The energy
   \be
   E=\int
   \frac{2}{(1+|W|^2)^2}(|\partial_t W|^2+|\partial_x W|^2+|\partial_y W|^2
   +m^2|W|^2)\,dxdy,
   \ee
   of the Q-lump solution (\ref{ratmap}) is infinite 
   if $\alpha_{N-1}\ne 0,$ because the field does not decay sufficiently rapidly for the integral to converge. In particular, this means that there are no finite energy Q-lumps with $N=1$. For $N>1$ there is a $(4N-2)$-dimensional moduli space $\widetilde{\cal M}_N$ of finite energy stationary Q-lumps given by setting $\alpha_{N-1}=0.$ This reflects the fact that there are no forces between the stationary Q-lumps given by these solutions, although there are some constraints relating the parameters of the individual Q-lump constituents.

   \section{Moduli space dynamics}\quad
   The dynamics of Q-lumps can be investigated by applying the moduli space approximation \cite{Ma} that is a cornerstone in the study of topological soliton dynamics, having been applied to investigate the dynamics of a wide variety of soliton systems. A slightly unusual feature in the application to Q-lumps is that the dynamics is restricted to motion on a moduli space of stationary soliton solutions, whereas the typical application is to approximate soliton evolution via restricting the motion to a moduli space of static soliton solutions. However, the basic principle remains the same and is implemented as follows.
   
   Let $q_i$, for $i=1,...,4N-2,$ be real coordinates on the moduli space $\widetilde{\cal M}_N$, given by the real and imaginary parts of $\alpha_0,...,\alpha_{N-2},\beta_0,...,\beta_{N-1}$ that appear in (\ref{ratmap}). Allowing these coordinates to be time-dependent, ${\bf q}(t)$, it is convenient to absorb the overall factor $e^{imt}$ for the stationary solutions into the time dependence of the moduli space coordinates. Explicitly, $W$ is approximated by the restricted form
   \be
   W(z;{\bf q}(t))=\frac{(q_{2N-3}+iq_{2N-2})z^{N-2} +\cdots +(q_3+iq_4)z+q_1+iq_2}{z^N+(q_{4N-3}+iq_{4N-2})z^{N-1} +\cdots +(q_{2N+1}+iq_{2N+2})z+q_{2N-1}+iq_{2N}}.
   \ee
    Substituting this form into the Lagrangian density (\ref{cp1lag}) and performing the integration over space yields the Lagrangian
   \be
   L=g_{ij}\dot q_i\dot q_j - V -4\pi N,
   \label{mslag}
   \ee
   where $\dot q_i=dq_i/dt$ and the metric is
   \be
   g_{ij}({\bf q})=\int \frac{1}{(1+|W|^2)^2}
   \bigg(
   \frac{\partial W}{\partial q_i}\frac{\partial \overline W}{\partial q_j}
   +   \frac{\partial W}{\partial q_j}\frac{\partial \overline W}{\partial q_i}
   \bigg)\,dxdy,
   \label{metric}
   \ee
   with the potential 
   \be
   V({\bf q})=2m^2\int \frac{|W|^2}{(1+|W|^2)^2}\,dxdy.
   \label{pot}
   \ee
   The equations of motion that follow from (\ref{mslag}) are
   \be
   2g_{ki}\ddot q_i+\bigg(2\frac{\partial g_{ki}}{\partial q_j}-
   \frac{\partial g_{ij}}{\partial q_k}\bigg)\dot q_i\dot q_j+\frac{\partial V}{\partial q_k}=0,
   \label{odes}
   \ee
   where geodesic motion is modified by the force due to the potential.
   
As there are no finite energy Q-lumps with $N=1$, the simplest case to consider is $N=2$, where motion takes place on the 6-dimensional manifold  $\widetilde{\cal M}_2$, with the associated field
   \be
   W=\frac{q_1+iq_2}{z^2+(q_5+iq_6)z+q_3+iq_4}.
   \label{ch2}
   \ee
   The centre of mass may be fixed at the origin by setting $q_5=q_6=0$, to
   yield a 4-dimensional submanifold $\widetilde{\cal M}_2^0$ of $\widetilde{\cal M}_2$, with coordinates $q_1,q_2,q_3,q_4.$ 
   The moduli space dynamics on the manifold $\widetilde{\cal M}_2^0$ was studied some time ago in great detail by Leese \cite{Le}. The remainder of this section provides a brief review of this work, to set the scene for the following sections where new results are presented.

   For $N=2$ the integrals required to calculate the metric (\ref{metric}) and the potential (\ref{pot}) can be evaluated in terms of elliptic integrals, although this will not be exploited here as the integrals will be computed numerically, in order to use the same methods that will be applied later for higher charges. The ordinary differential equations in (\ref{odes}) are solved numerically using a variable stepsize Runge-Kutta method, with the initial conditions provided as follows. 
   Setting
   \be
   q_1=2\Lambda(A\cos\Theta-B\sin\Theta),\
   q_2=2\Lambda(A\sin\Theta+B\cos\Theta),\
   q_3=B^2-A^2,\
   q_4=-2AB,
   \ee
   allows the field (\ref{ch2}) to be rewritten as
   \be
   W=\bigg(\frac{\Lambda}{z-A-iB}-\frac{\Lambda}{z+A+iB}\bigg)e^{i\Theta},
   \label{2well}
   \ee
   providing an interpretation of the parameters in terms of a pair of well-separated unit charge Q-lumps with equal size $\Lambda$ and positions in the $(x,y)$-plane given by $(A,B)$ and $(-A,-B)$, valid for $A^2+B^2\gg \Lambda^2.$ Note that the pair of Q-lumps have a relative phase angle of $\pi$, because of the minus sign between the two terms in (\ref{2well}).
The fact that both Q-lumps have the same size, and their relative phase is locked at $\pi$, is a consequence of the finite energy constraint in this case.
   The scattering of a pair of Q-lumps, with an initial motion parallel to the $x$-axis, can therefore be studied using the initial conditions
   \be
   \Theta(0)=0,\
   \dot\Theta(0)=m,\
   \Lambda(0)=\lambda,\
   \dot\Lambda(0)=0,\
   A(0)=a,\
   \dot A(0)=v,\
   B(0)=b,\
   \dot B(0)=0,
   \label{ic}
   \ee
   where $\lambda$ is the initial common size, $\pm(a,b)$ are the initial positions and $v$ is the initial common speed of each Q-lump.
\begin{figure}[ht]\begin{center}
    \includegraphics[width=0.8\columnwidth]{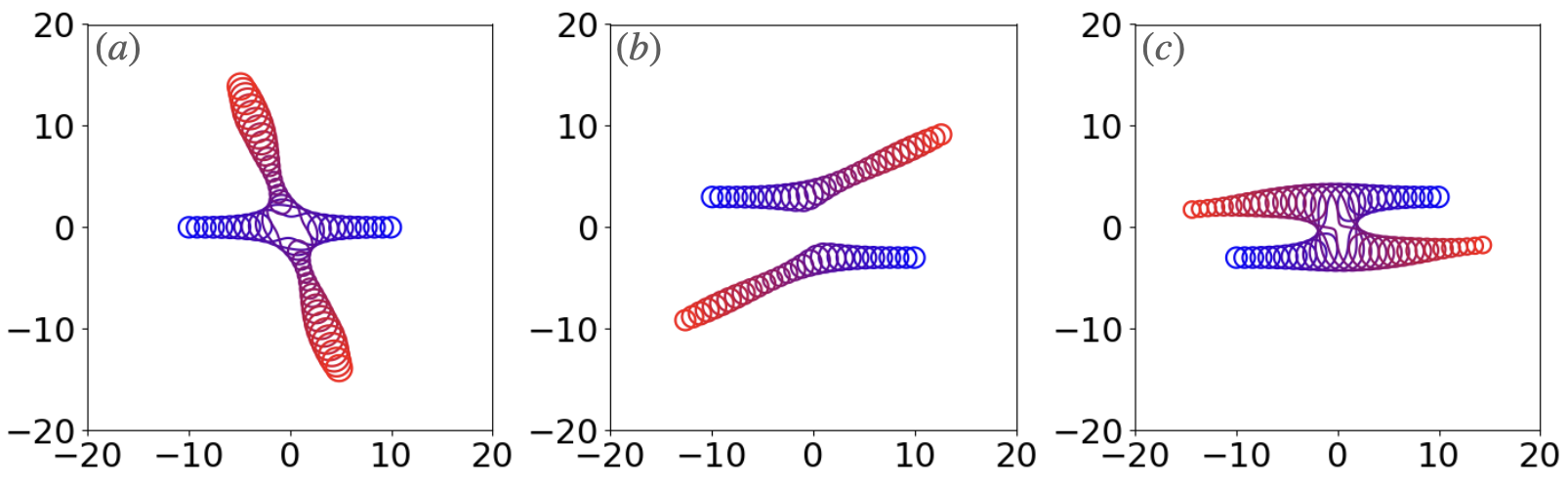}
          \caption{Moduli space dynamics for $N=2$ for times $t\in[0,120]$ using the parameters $\lambda=1,\,a=-10,\,v=0.2$ with (a) $b=0$; (b) $b=3$; (c) $b=-3$.}
          \label{mex2}\end{center}\end{figure}

Without loss of generality, the initial size may be set to $\lambda=1$. The initial separation in the $x$-direction is fixed by choosing $a=-10$, so that the Q-lumps are reasonably well-separated at the start of the motion. To present the resulting evolution, a variant of the multiple exposure plotting technique introduced in \cite{Le} will be applied. This involves plotting, in a single image, the curves given by the level set $\phi_3=0$ (or equivalently $|W|=1$) at equally spaced time intervals. The curves are coloured blue in the early stages of the motion and transition to red as time increases.

Fig.\ref{mex2}(a) displays a typical head-on scattering process ($b=0$) with initial speed $v=0.2$. The Q-lump that begins at the left ends up at the top of the image, allowing the identification of a scattering angle between the initial and final directions of motion of around $2\pi/3.$ This agrees with the results found in \cite{Le}, where the scattering angle as a function of initial speed $v$ is found to lie in the interval $(\pi/2,\pi)$, with a value of around $2\pi/3$ for a wide range of $v$. The scattering angle is a monotonically decreasing function of $v$, tending towards $\pi/2$ in the fast scattering limit, where the interaction time is much shorter than the period of the internal rotation. This is expected, as the traditional $\pi/2$ scattering of a range of topological solitons, including lumps in the pure $\sigma$-model, should be recovered in this limit. The collision induces an oscillation of the size of the Q-lumps that is clearly visible in the multiple exposure plot. The amplitude of this oscillation decreases if the initial scattering speed $v$ is decreased.

Scattering with a positive impact parameter $b=3$ is displayed in Fig.\ref{mex2}(b). The results for $b>0$ are not surprising, with the scattering angle being a monotonically decreasing function of $b$, tending to zero in the large $b$ limit. Scattering is not symmetric under the replacement $b\to -b$, as illustrated in Fig.\ref{mex2}(c) by the example with $b=-3$. This lack of symmetry is due to the fact that the internal phase rotation is anti-clockwise and therefore selects a preferred sign for comparison with the angular momentum generated by the collision. The combination $b\to -b$ together with $m\to -m$ is required to generate a symmetric scattering event. For $b<0$ the scattering angle is a complicated function of both $b$ and $v$, with regions with a highly sensitive dependence, as the Q-lumps can orbit around each other many times before they eventually escape, or merge to lose their individual identities before separating \cite{Le}.
  
 \section{Cyclic scattering}\quad
 For $N>2$ the study of generic finite energy Q-lump scattering would be rather cumbersome using the moduli space approximation, because $\widetilde{\cal M}_N$ has dimension $4N-2$. Fixing a centre of mass reduces the dimension to $4(N-1)$, but this is still a little unwieldy. To avoid this difficulty, in this section the dynamics on moduli space will be investigated on various 4-dimensional submanifolds of $\widetilde{\cal M}_N$, obtained by imposing cyclic symmetries.

Rather than thinking of fixing the centre of mass to obtain the 4-dimensional submanfold $\widetilde{\cal M}_2^0$ of $\widetilde{\cal M}_2$, an alternative point of view to obtain this submanifold is to impose the cyclic $C_2$ symmetry $W(-z)=W(z)$, given by a spatial rotation around the origin by $\pi$. The fact that both Q-lumps have the same size, and their relative phase is locked at $\pi$, is then a clear consequence of the $C_2$ symmetry that exchanges the pair of Q-lumps. The generalization to study Q-lump scattering with charge $N$ and cyclic $C_N$ symmetry is therefore rather natural. 

Let $\omega_N=e^{2\pi i/N}$ denote the $N^{\rm th}$ root of unity and define $\Sigma_N^j$, for $j=0,...,N-2$, to be the 4-dimensional submanifold of $\widetilde{\cal M}_N$ obtained by imposing the cyclic $C_N$ symmetry $W(\omega_Nz)=\omega_N^jW(z)$. The corresponding field is given by
\be
W=\frac{z^j(q_1+iq_2)}{z^N+q_3+iq_4},
\label{cyclic}
\ee
where a convenient relabelling of the indices of ${\bf q}$ has been applied so that the coordinates on this submanifold are $q_1,q_2,q_3,q_4$.
Note that the submanifold $\Sigma_N^0$ contains the axially symmetric Q-lump with charge $N$, given by $W=(q_1+iq_2)/z^N$, but this is not contained in any of the other submanifolds, $\Sigma_N^j$ with $j>0$.

In the asymptotic region of well-separated unit charge Q-lumps, the field (\ref{cyclic}) may be rewritten
as
\be
W=e^{i\Theta}\Lambda \sum_{k=0}^{N-1}\frac{\omega_N^{(j+1)k}}{z-(A+iB)\omega_N^k},
\label{asig}
\ee
revealing $N$ unit charge Q-lumps of equal size $\Lambda$ on the vertices of a regular $N$-gon, with a frozen relative phase of
$2(j+1)\pi/N$ between neighbouring Q-lumps. The scattering discussed in the previous section corresponds to the case $N=2$, where there is only one submanifold, $\Sigma_2^0$, and the relative phase is frozen at $\pi$.

\begin{figure}[!ht]\begin{center}
    \includegraphics[width=0.5\columnwidth]{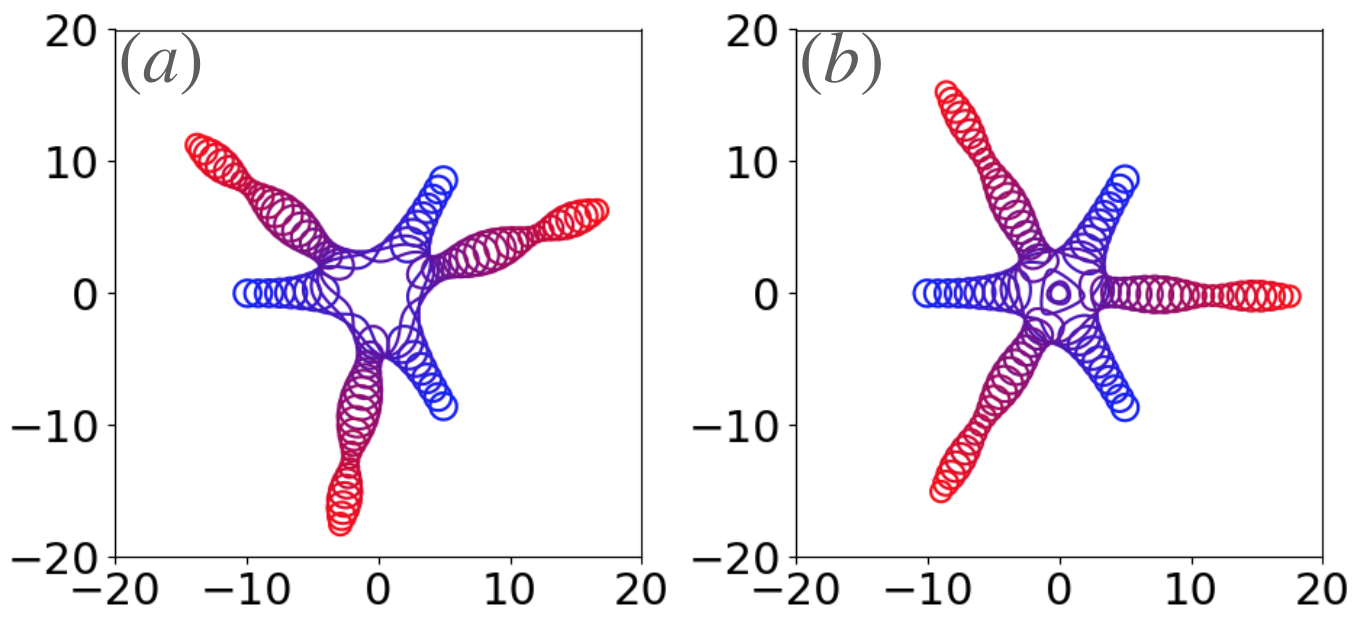}
          \caption{Moduli space dynamics for $N=3$ with $C_3$ symmetry for times $t\in[0,260]$ using $\lambda=1,\,a=-10,\,b=0,\,v=0.1$ on (a) $\Sigma_3^0$; (b) $\Sigma_3^1$.}
          \label{mex3c3}\end{center}\end{figure}

These submanifolds therefore allow the first studies of Q-lump dynamics where relative phases are not frozen at $\pi$. The asymptotic formula (\ref{asig}) is a clear generalization of the $N=2$ formula (\ref{2well}), and the same initial conditions (\ref{ic}) can be used for the parameter values.
Fig.\ref{mex3c3}(a) illustrates the scattering of three Q-lumps with $C_3$ symmetry and relative phases of $2\pi/3$ between the Q-lumps, that is, dynamics on the submanifold $\Sigma_3^0.$ The scattering angle lies in the interval $(2\pi/3,\pi)$, and tends towards the lower limit of this interval as the initial speed increases. This is the expected generalization of the scattering on $\Sigma_2^0$, for the following reason.

In the case of cyclic $C_N$ scattering of $N$ lumps in the pure $\sigma$-model, the similar scattering process results in lumps on the vertices of an outgoing regular $N$-gon that is the dual of the incoming $N$-gon \cite{KPZ}. This has been termed $\pi/N$ scattering, where the scattering angle refers to the rotation of the polygon, because the lumps lose their individual identities during the scattering process so a scattering angle for any individual lump is ill-defined. However, in the case of Q-lumps, the individual Q-lumps remain distinct enough to define the scattering angle of a single Q-lump, as earlier, and translating the polygon scattering into this definition yields a scattering angle of $\pi (N-1)/N.$ Hence the expected lower limit of $2\pi/3$ in the above case of $\Sigma_3^0$, where $N=3.$

Fig.\ref{mex3c3}(b) displays the equivalent scattering process to Fig.\ref{mex3c3}(a), with the same parameter values, but now on the submanifold $\Sigma_3^1$, so the only change is that the relative phase between the Q-lumps is now $4\pi/3$ rather than $2\pi/3$. The change in relative phase has reduced the scattering angle, which is now slightly less than $2\pi/3$, and is therefore outside the interval found for scattering on $\Sigma_3^0.$ There is also a slight reduction in the amplitude of the size oscillations of the outgoing Q-lumps. Note that the configuration formed at the point of closest approach is very different on $\Sigma_3^0$ and $\Sigma_3^1$, reflecting the fact that only the former submanifold contains the axially symmetric charge three Q-lump.
\begin{figure}[!hb]\begin{center}
  \includegraphics[width=0.9\columnwidth]{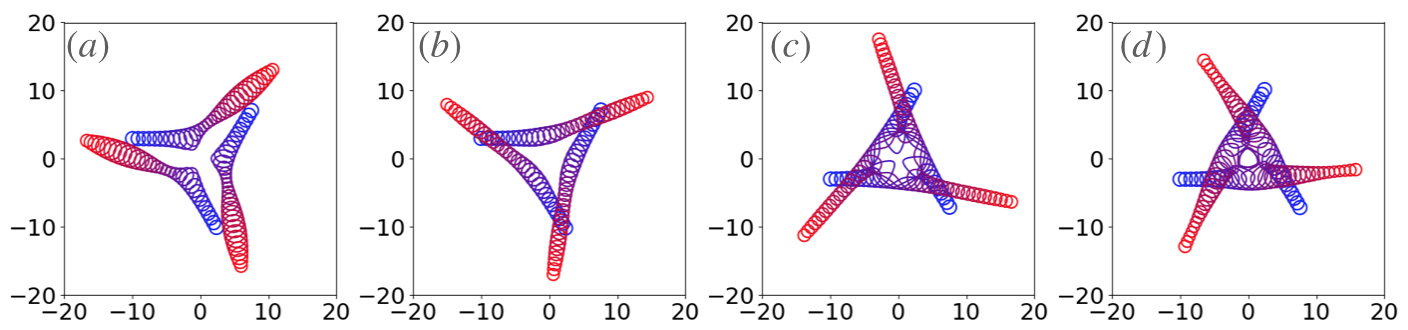}
   \caption{Moduli space dynamics for $N=3$ with $C_3$ symmetry for times $t\in[0,130]$ using $\lambda=1,\,a=-10,\,v=0.2$ on (a) $\Sigma_3^0$ with $b=3$; (b) $\Sigma_3^1$ with $b=3$; (c) $\Sigma_3^0$ with $b=-3$; (d) $\Sigma_3^1$ with $b=-3$.}
   \label{mex3c3b3}\end{center}\end{figure}

Scatterings on the submanifolds $\Sigma_3^0$ and $\Sigma_3^1$, with both positive and negative impact parameters, can be found in Fig.\ref{mex3c3b3}. These results show similar features to the charge two case, with the scattering angle again reduced on $\Sigma_3^1$ compared to $\Sigma_3^0$.

\begin{figure}[!ht]\begin{center}
    \includegraphics[width=0.8\columnwidth]{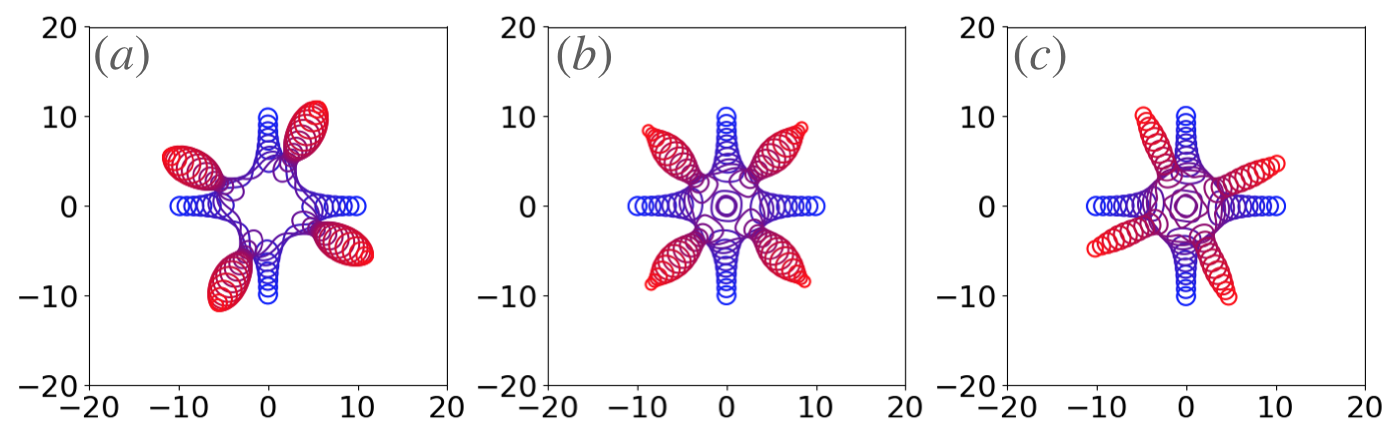}
          \caption{Moduli space dynamics for $N=4$ with $C_4$ symmetry for times $t\in[0,200]$ using $\lambda=1,\,a=-10,\,b=0,\,v=0.1$ on (a) $\Sigma_4^0$; (b) $\Sigma_4^1$; (c) $\Sigma_4^2$.}
          \label{mex4c4}\end{center}\end{figure}

Charge four scattering with $C_4$ symmetry provides access to three different sets of relative phases. Examples on the submanifolds $\Sigma_4^0,\Sigma_4^1,\Sigma_4^2$ are displayed in Fig.\ref{mex4c4}, where the same parameter values are used in each case. This provides further evidence that increasing the relative phase between neighbouring Q-lumps reduces both the scattering angle and the amplitude of the size oscillation of the outgoing Q-lumps.

So far, all the examples studied have involved Q-lumps that all have the same size. It is possible to gain access to scattering events that involve Q-lumps with different sizes by considering another family of 4-dimensional submanifolds, this time by imposing cyclic $C_{N-1}$ symmetry in the charge $N$ sector.
For $N>2$, define the 4-dimensional submanifold, $\Xi_N$, of $\widetilde{\cal M}_N$ by imposing the cyclic $C_{N-1}$ symmetry $W(\omega_{N-1}z)=\omega_{N-1}^{N-2}W(z)$. In this case the field takes the form
\be
W=\frac{q_1+iq_2}{z^N+z(q_3+iq_4)},
\ee
that includes the axially symmetric Q-lump with charge $N$.
In the asymptotic region of well-separated Q-lumps, the field may be rewritten
as
\be
W=e^{i\Theta}\Lambda\bigg(\frac{1-N}{z}+ \sum_{k=0}^{N-2}\frac{1}{z-(A+iB)\omega_{N-1}^k}\bigg),
\ee
revealing $N-1$ unit charge Q-lumps with size $\Lambda$ on the vertices of a regular $(N-1)$-gon and a unit charge Q-lump at the origin with size $(N-1)\Lambda$. All the Q-lumps on the $(N-1)$-gon have the same phase, but there is a phase difference of $\pi$ between any of these Q-lumps and the Q-lump at the origin.

\begin{figure}[ht]\begin{center}
    \includegraphics[width=0.6\columnwidth]{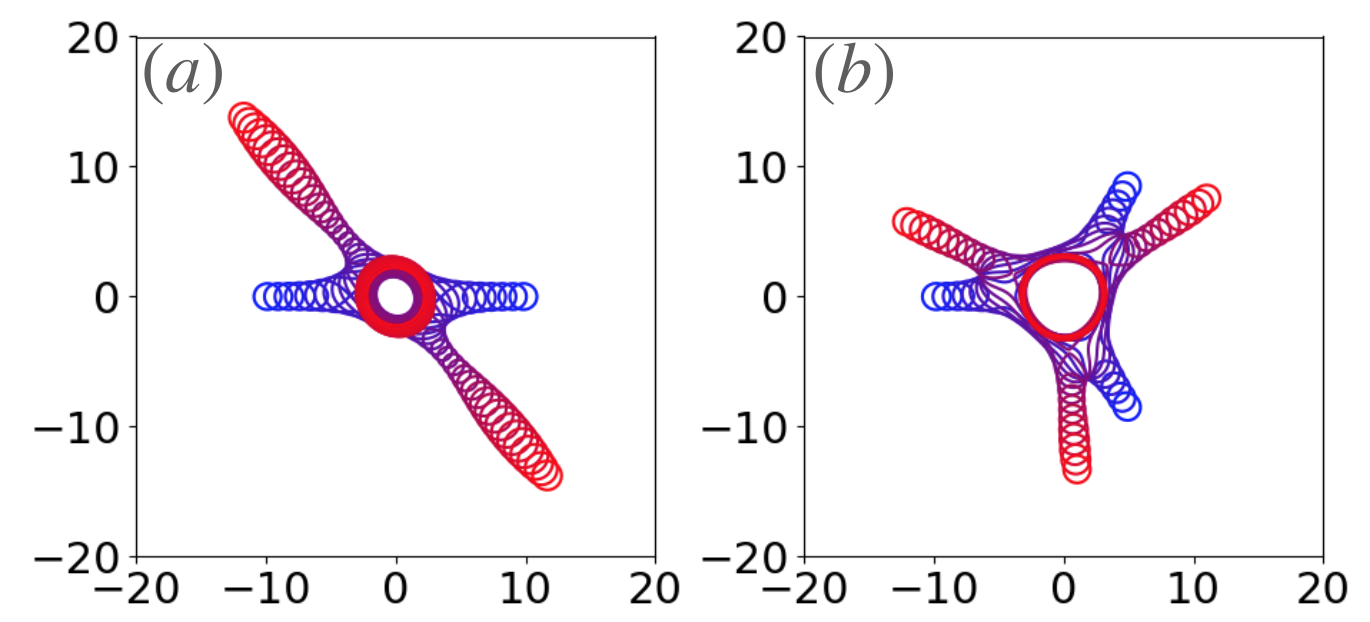}
          \caption{Moduli space dynamics with $N-1$ unit charge Q-lumps scattering on a larger unit charge Q-lump at the origin, using $\lambda=1,\,a=-10,\,b=0,\,v=0.1$ for (a) $N=3$ with $C_2$ symmetry on $\Xi_3$; (b) $N=4$ with $C_3$ symmetry on $\Xi_4$.}
          \label{mex3c24c3}\end{center}\end{figure}

Fig.\ref{mex3c24c3} presents examples on the submanifolds $\Xi_3$ and $\Xi_4$. The larger Q-lump remains at the origin, with a visible distortion of the axial symmetry to $C_{N-1}$ symmetry, and its size oscillation is synchronized with the oscillation of the smaller Q-lumps, as it must to keep the ratio of the sizes equal to $N-1$. A comparison of Fig.\ref{mex3c24c3}(b) and Fig.\ref{mex3c3}(a) reveals that the presence of the central Q-lump increases the scattering angle and reduces the amplitude of the oscillation in size.

\begin{figure}[!hb]\begin{center}
    \includegraphics[width=0.6\columnwidth]{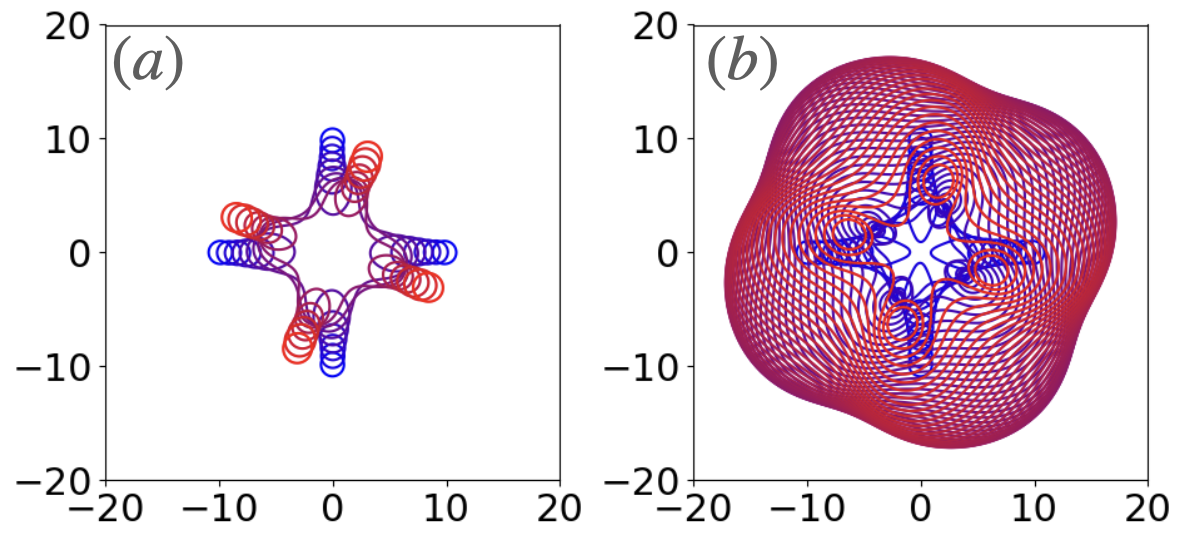}
    \caption{Moduli space dynamics for $N=4$ with $C_4$ symmetry on the submanifold  $\Sigma_4^0$ using $\lambda=1,\,a=-10,\,b=0,$ with (a) $v=0.05$; (b) $v=0.2$.}
          \label{speeds}\end{center}\end{figure}

As mentioned earlier, a general feature of Q-lump scattering is that an increase in the speed of the incoming Q-lumps produces an increase in the amplitude of the size oscillation of the outgoing Q-lumps. An interesting example to investigate the consequences of varying the initial speed is to take the $\Sigma_4^0$ scattering shown in Fig.\ref{mex4c4}(a) and vary the initial speed from $v=0.1$. Reducing the initial speed to $v=0.05$ results in the scattering presented in Fig.\ref{speeds}(a), with a clear reduction in the amplitude of the size oscillation, as there is less kinetic energy to transfer to this mode. A much more interesting phenomenon is found by increasing the speed to $v=0.2$, as displayed in Fig.\ref{speeds}(b). The scattering process is now more complicated, making it a little difficult to decipher the information contained within the multiple exposure plot. However, it is clear that the scattering is now qualitatively different from the the previous examples. An analysis of this scattering event will be presented later, with a series of energy density plots from full field simulations
being easier to interpret than this single multiple exposure plot.

\section{Field theory dynamics}\quad
The $\bphi$ field formulation is used to perform full field simulations of Q-lump dynamics. The field equation that follows from the variation of (\ref{lag}) is
\be
\partial_\mu\partial^\mu\bphi+(\partial_\mu\bphi\cdot\partial^\mu\bphi)\bphi+m^2\phi_3(\phi_3\bphi-{\bf e}_3)=0,
\ee
where ${\bf e}_3=(0,0,1).$ This nonlinear partial differential equation is solved numerically on a square lattice consisting of $1501\times1501$ lattice points, with lattice spacings $\Delta x=\Delta y=0.04$, to give a spatial simulation region of $[-30,30]\times[-30,30]$. Spatial derivatives are computed using a fourth-order finite difference approximation and time evolution is performed using a fourth-order Runge-Kutta scheme with a fixed timestep $\Delta t=0.01$.

The boundary condition at the edge of the simulation lattice is taken to be compatible with the internal rotation of a Q-lump, in either direction, by imposing the evolution equation $\partial_t\partial_t W=-m^2 W$. In terms of the $\bphi$ field this becomes
\bea
&\partial_t\partial_t\phi_1&-2\frac{\partial_t\phi_1\partial_t\phi_3}{1+\phi_3}
+\phi_1\bigg(\frac{|\partial_t\bphi|^2}{1+\phi_3}+m^2\phi_3\bigg)=0\\
&\partial_t\partial_t\phi_2&-2\frac{\partial_t\phi_2\partial_t\phi_3}{1+\phi_3}
+\phi_2\bigg(\frac{|\partial_t\bphi|^2}{1+\phi_3}+m^2\phi_3\bigg)=0\\
&\partial_t\partial_t\phi_3&-2\frac{\partial_t\phi_3\partial_t\phi_3}{1+\phi_3}
+|\partial_t\bphi|^2-m^2(1-\phi_3^2)=0.
\eea

\begin{figure}[!hb]\begin{center}
    \includegraphics[width=1.0\columnwidth]{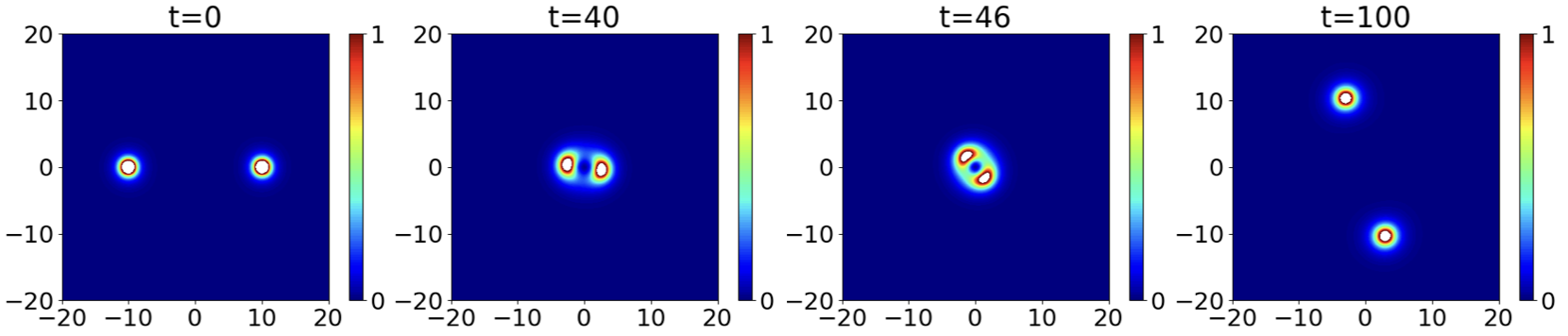}
    \caption{
      Energy density plots from field theory dynamics that reproduces the $N=2$ scattering on $\Sigma_2^0$ shown in Fig.\ref{mex2}(a).}
          \label{2b0}\end{center}\end{figure}

\begin{figure}[ht]\begin{center}
    \includegraphics[width=1.0\columnwidth]{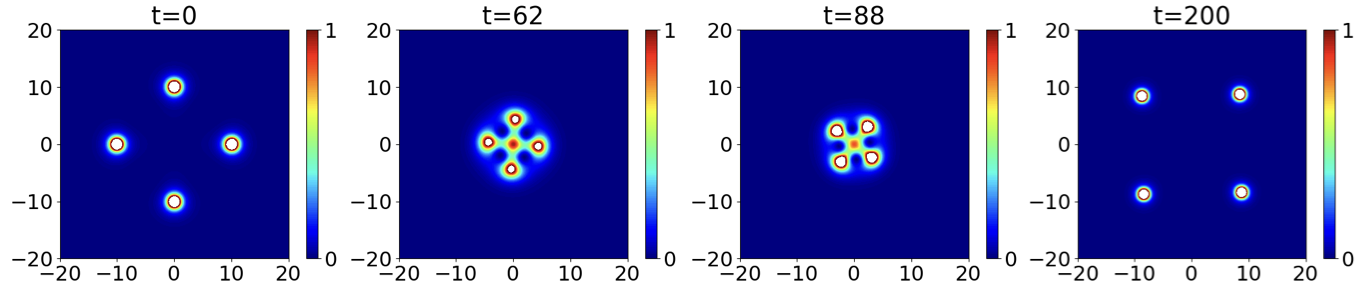}
    \caption{
Energy density plots from field theory dynamics that reproduces the cyclic $N=4$ scattering on $\Sigma_4^1$ shown in Fig.\ref{mex4c4}(b).}
          \label{sig41}\end{center}\end{figure}

To check the results of the moduli space approximation, the initial conditions for the field theory simulations, $\bphi|_{t=0}$ and $\partial_t\bphi|_{t=0}$, are taken to be the same as in the moduli space dynamics. It is found that the examples presented in the previous sections are in excellent agreement with the field theory simulations, thereby providing a good cross-check on both methods.  Representative examples are provided in Fig.\ref{2b0}, which reproduces the charge two scattering on $\Sigma_2^0$ found in Fig.\ref{mex2}(a), and Fig.\ref{sig41}, which reproduces the cyclic charge four scattering on $\Sigma_4^1$ from Fig.\ref{mex4c4}(b).

\begin{figure}[!ht]\begin{center}
    \includegraphics[width=0.5\columnwidth]{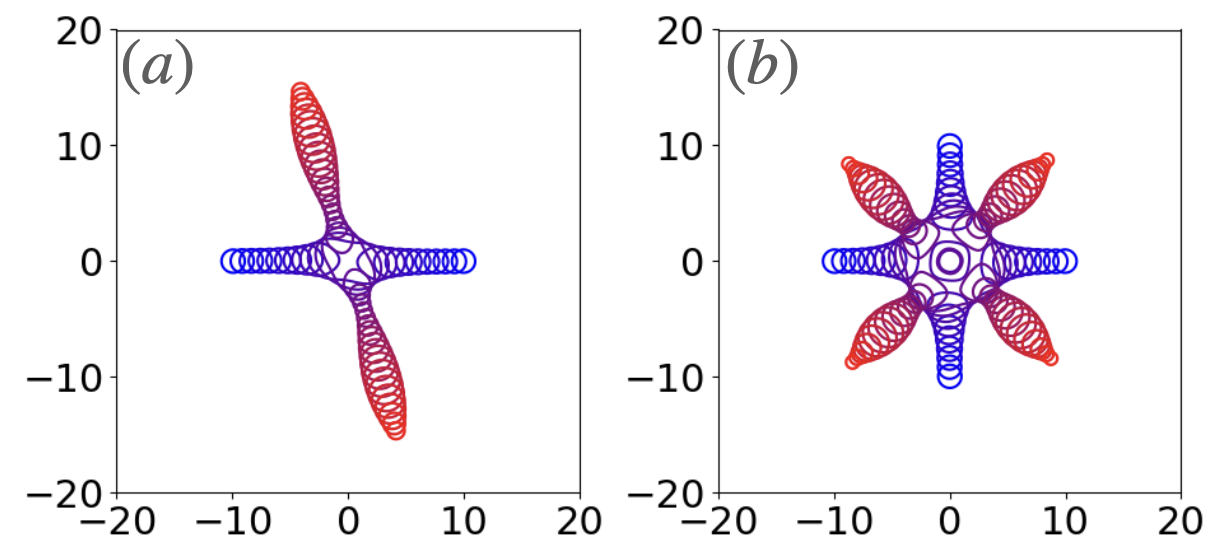}
          \caption{Field theory simulations represented as multiple exposure plots for comparison with moduli space dynamics, (a) compare with Fig.\ref{mex2}(a); (b) compare with Fig.\ref{mex4c4}(b).}
          \label{check}\end{center}\end{figure}

In these figures the energy density at various times is visualized using a heat map, with the colour bar provided for values in the interval $[0,1]$ and values greater than one displayed as white. For clarity, only the region $[-20,20]\times[-20,20]$ of the full simulation domain $[-30,30]\times[-30,30]$ is shown. These two simulations are also available as short movies, see the supplementary data files m01.mp4 and m02.mp4. To aid comparison with the results from moduli space dynamics, these simulations are also presented as multiple exposure plots, see Fig.\ref{check}(a) (to be compared with Fig.\ref{mex2}(a)) and Fig.\ref{check}(b) (to be compared with Fig.\ref{mex4c4}(b)). This makes the excellent agreement obvious.

\begin{figure}[!ht]\begin{center}
    \includegraphics[width=1.0\columnwidth]{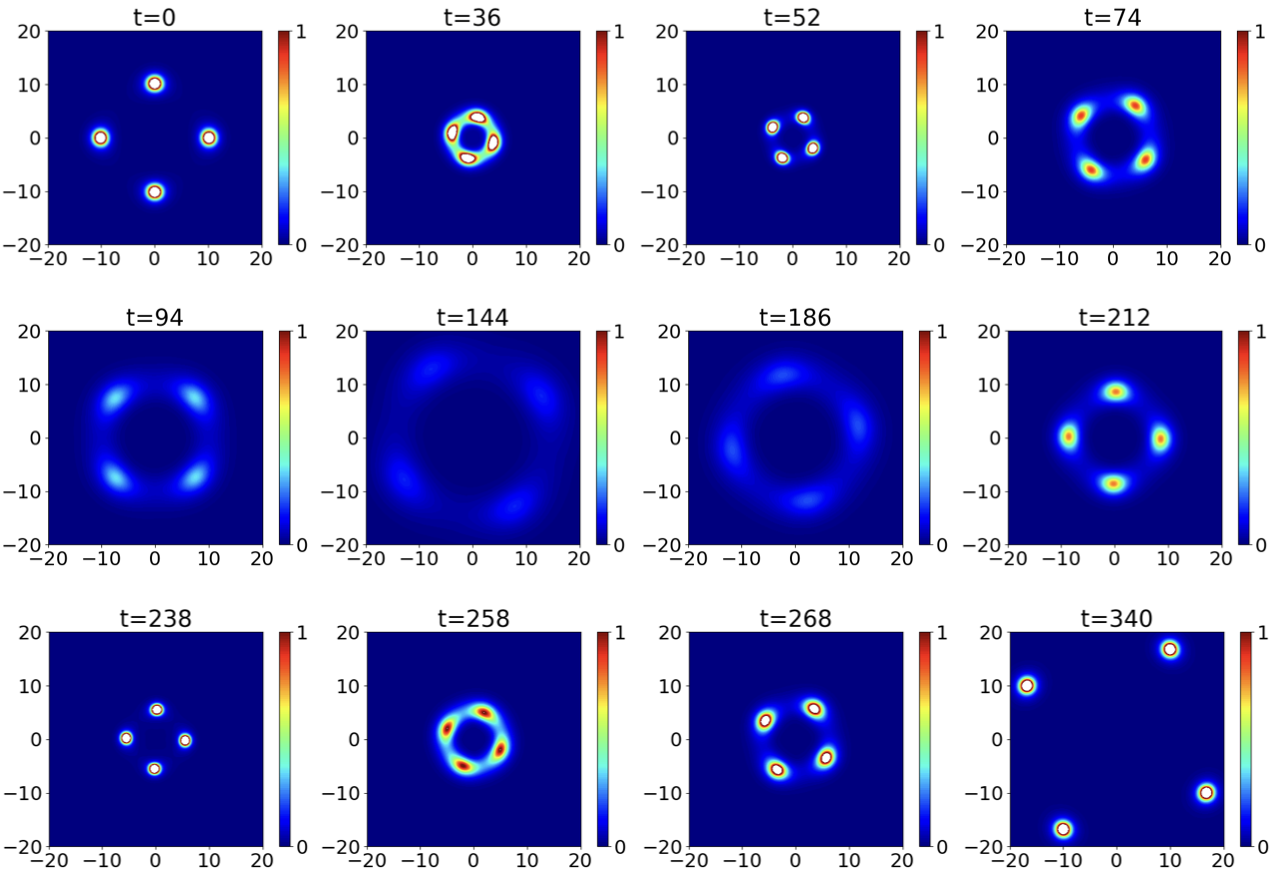}
          \caption{Energy density plots from field theory dynamics associated with the cyclic charge four scattering of $\Sigma_4^0$, with an initial speed $v=0.2$.}
          \label{ensig41}\end{center}\end{figure}

It is now time to return to the exotic scattering event presented at the end of the previous section in Fig.\ref{speeds}(b), to investigate this process via full field simulations. The resulting energy density plots are displayed in Fig.\ref{ensig41}, for times that include and go beyond the multiple exposure plot from moduli space dynamics that is shown in Fig.\ref{speeds}(b). This simulation is also available as the movie m03.mp4 in the supplementary data. Initially the scattering is similar to that found at lower speeds, but a difference at this higher speed can already be seen at $t=74$, where the expansion of the size of the Q-lumps is comparable to their increase in separation. Q-lumps are well-separated if their separation is large compared to their size, but if their size grows sufficiently rapidly then they are not well-separated even as they move apart. This is the situation found at $t=94$, where the configuration resembles the ring of an axially symmetric charge four Q-lump with a slight square perturbation, rather than four distinct Q-lumps. The ring reaches a maximum size at $t=144$ and then the size oscillation enters the phase of size reduction that shrinks the ring, as the small distinct Q-lumps reappear at $t=238$. The Q-lumps now scatter again, in a similar manner to the early stage of the scattering, but this time the size oscillation is not sufficient to prevent well-separated Q-lumps from emerging and escaping. This exotic double scattering process is made possible at high speeds by a sufficient transfer of kinetic energy into the size oscillation mode.

The scattering on $\Sigma_4^1$, shown in Fig.\ref{sig41} for $v=0.1$, also transforms to a similar complicated scattering if the initial speed is increased to $v=0.2$, but such a transformation is not found on the submanifold $\Sigma_4^2$ at the same speed $v=0.2$. This is consistent with the results presented in Fig.\ref{mex4c4} for $v=0.1$, where the amplitude of the size oscillation generated on $\Sigma_4^2$ is much less than on $\Sigma_4^0$ or $\Sigma_4^1$. Clearly, relative phases play an important role in the transfer of energy between different modes. The double scattering process, where soliton kinetic energy is transferred to another mode and then back to soliton kinetic energy, is reminiscent of the scattering of kinks and anti-kinks in some (1+1)-dimensional systems, which leads to resonant scattering and a fractal structure in the dependence of the final state on the initial collision speed \cite{CSW}. It might be interesting to investigate whether similar phenomena are possible in this (2+1)-dimensional soliton system.

Some scattering events in the charge four sector will now be considered, where the initial conditions consist of a pair of well-separated axially symmetric charge two Q-lumps. One of the differences in studying the scattering of a pair of charge two Q-lumps, rather than a pair of charge one Q-lumps, is that the finite energy constraint now allows the pair of Q-lumps to have different sizes and any value of the relative phase. Recall that a pair of charge one Q-lumps must have equal size and be phase locked with a relative phase of $\pi$, in order to have finite energy. In principle, this charge four scattering could be investigated using moduli space dynamics, but the moduli space $\widetilde{\cal M}_4$ is 14-dimensional, so even after fixing the centre of mass, the motion on a 12-dimensional space is still a little cumbersome. Therefore, the investigation will proceed via field theory simulations.

The initial conditions can be taken from the moduli space approximation, namely the initial fields $\bphi|_{t=0}$ and $\partial_t\bphi|_{t=0}$ are taken to agree with those obtained from the field
\be
W=\bigg(\frac{\lambda_1^2}{(z-a-vt-ib)^2}+\frac{e^{i\chi}\lambda_2^2}{(z+a+vt+ib)^2}\bigg)e^{imt}.
\label{ic4}
\ee
This describes a pair of axially symmetric charge two Q-lumps at positions $(x,y)=\pm (a,b)$, with initial sizes $\lambda_1$ and $\lambda_2$, and an initial relative phase $\chi.$ The Q-lumps are initially moving parallel to the $x$-axis, in opposite directions, with equal speed $v$.

\begin{figure}[!hb]\begin{center}
    \includegraphics[width=0.93\columnwidth]{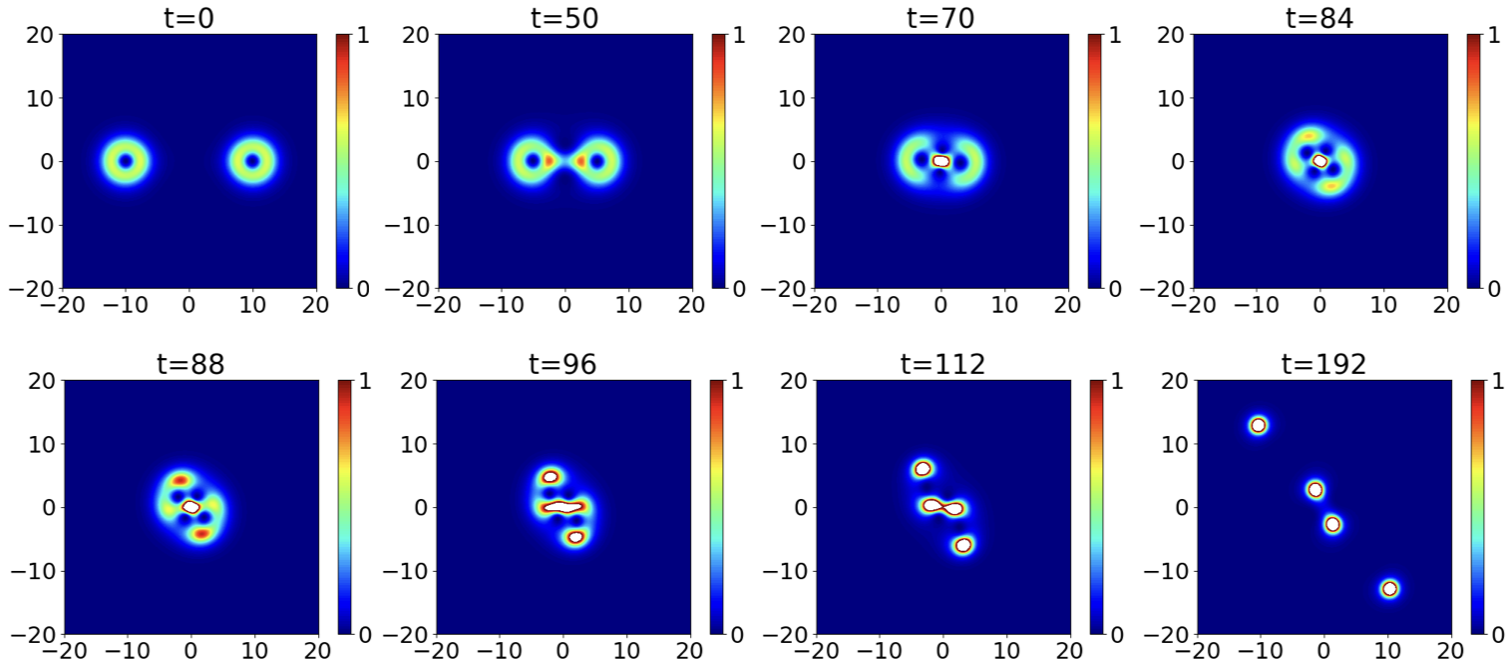}
    \caption{
Energy density plots from field theory dynamics for the scattering of a pair of charge two Q-lumps with parameters $\lambda_1=\lambda_2=3, \ a=-10, \ b=0, \ \chi=\pi, \ v=0.1.$}
          \label{4b0}\end{center}\end{figure}

To compare with the earlier scattering of a pair of charge one Q-lumps, the first simulations will consider equal sizes and a relative phase $\chi=\pi.$ An example of a head-on scattering ($b=0$) with $\lambda_1=\lambda_2=3$, and parameters $a=-10$ and $v=0.1$, is presented in Fig.\ref{4b0}.

\begin{figure}[ht]\begin{center}
    \includegraphics[width=0.73\columnwidth]{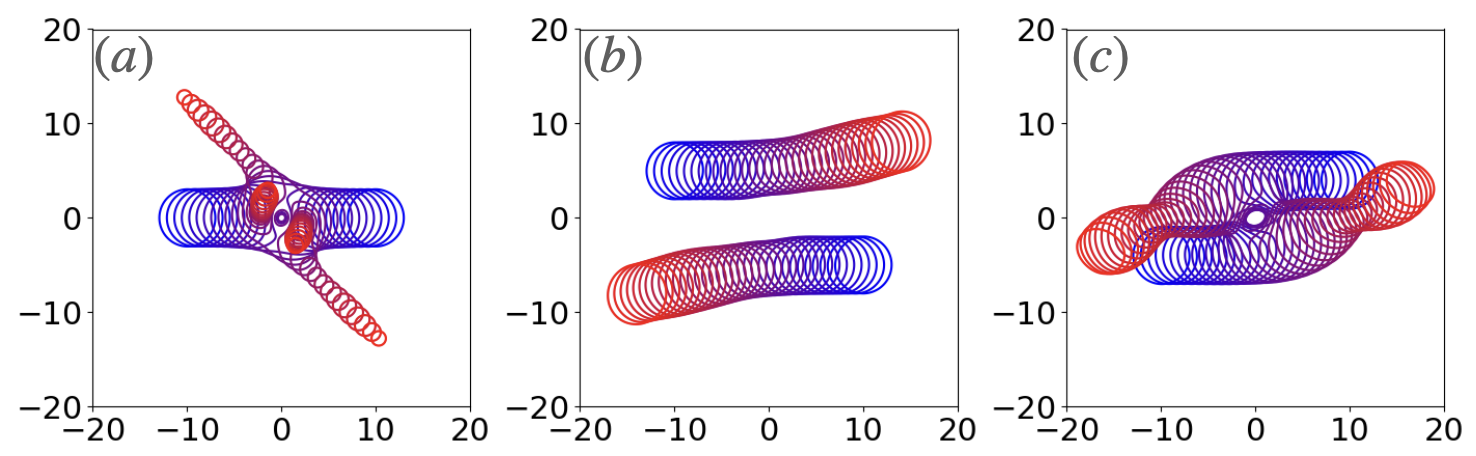}
    \caption{Field theory simulations of the scattering of a pair of charge two Q-lumps with $\lambda_1=\lambda_2=3, \ a=-10, \ \chi=\pi, \ v=0.1, \ $
      (a) $b=0$; (b) $b=5$; (c) $b=-4$. }
          \label{ex4b054}\end{center}\end{figure}

\begin{figure}[!hb]\begin{center}
    \includegraphics[width=0.85\columnwidth]{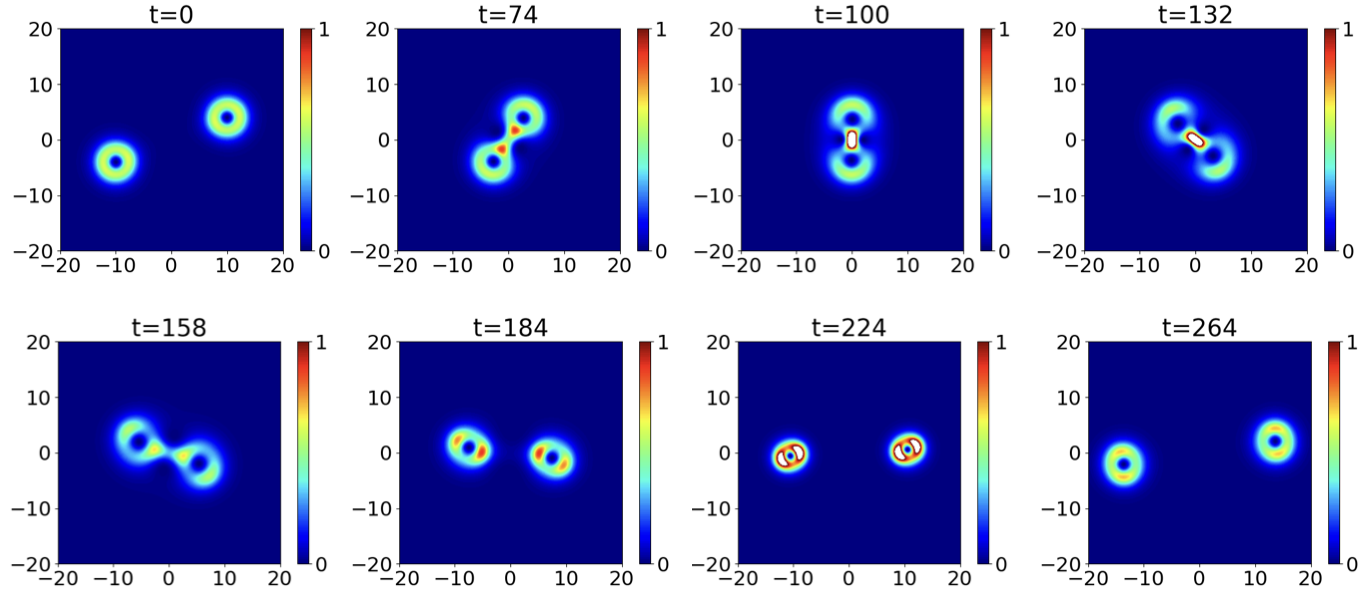}
    \caption{Energy density plots from field theory dynamics for the scattering of a pair of charge two Q-lumps with parameters $\lambda_1=\lambda_2=3, \ a=-10, \ b=-4, \ \chi=\pi, \ v=0.1.$}
    \label{4b4}\end{center}\end{figure}

The energy density plots in Fig.\ref{4b0}, and the corresponding movie m04.mp4, show the formation of a merged configuration at $t=84$, followed by the fission into four distinct Q-lumps. Two of the Q-lumps remain close to the origin, while the other two carry most of the kinetic energy as they move away. This scattering is presented as a multiple exposure plot in Fig.\ref{ex4b054}(a).
The result of introducing a non-zero impact parameter, $b$, is presented in Fig.\ref{ex4b054}(b) for $b=5$, and in Fig.\ref{ex4b054}(c) for $b=-4$. This mirrors the behaviour found earlier for scattering in the charge two sector. For the positive impact parameter the Q-lumps are deflected away from each other, with the scattering producing very little deformation to the Q-lumps. For the negative impact parameter the Q-lumps are attracted towards each other and the scattering induces a more significant deformation of the Q-lumps, but not enough to fission the charge two Q-lumps into individual charge one Q-lumps. The deformation of the outgoing Q-lumps is more clearly visible in the energy density plots in Fig.\ref{4b4} and the associated movie m05.mp4. 

\begin{figure}[ht]\begin{center}
    \includegraphics[width=0.7\columnwidth]{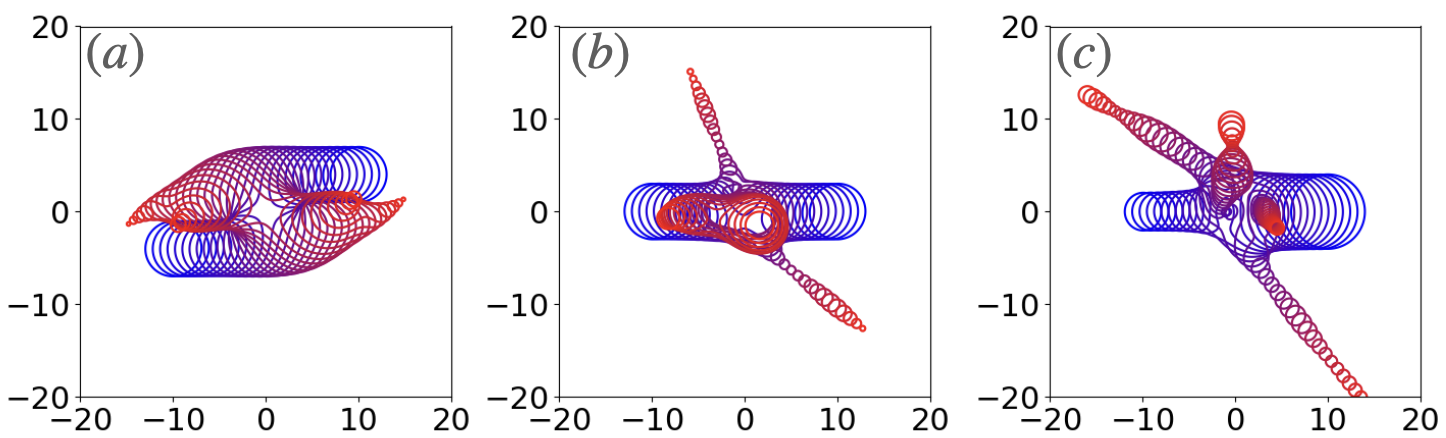}
    \caption{Field theory simulations of the scattering of a pair of charge two Q-lumps with $a=-10, \ v=0.1$ and remaining parameters,
      (a) $\lambda_1=\lambda_2=3, \ b=-4, \ \chi=0$;
      (b) $\lambda_1=\lambda_2=3, \ b=0, \ \chi=\pi/2$;
      (c) $\lambda_1=2,\ \lambda_2=4, \ b=0, \ \chi=\pi$.
      }
        \label{sb22}\end{center}\end{figure}

\begin{figure}[!hb]\begin{center}
    \includegraphics[width=0.8\columnwidth]{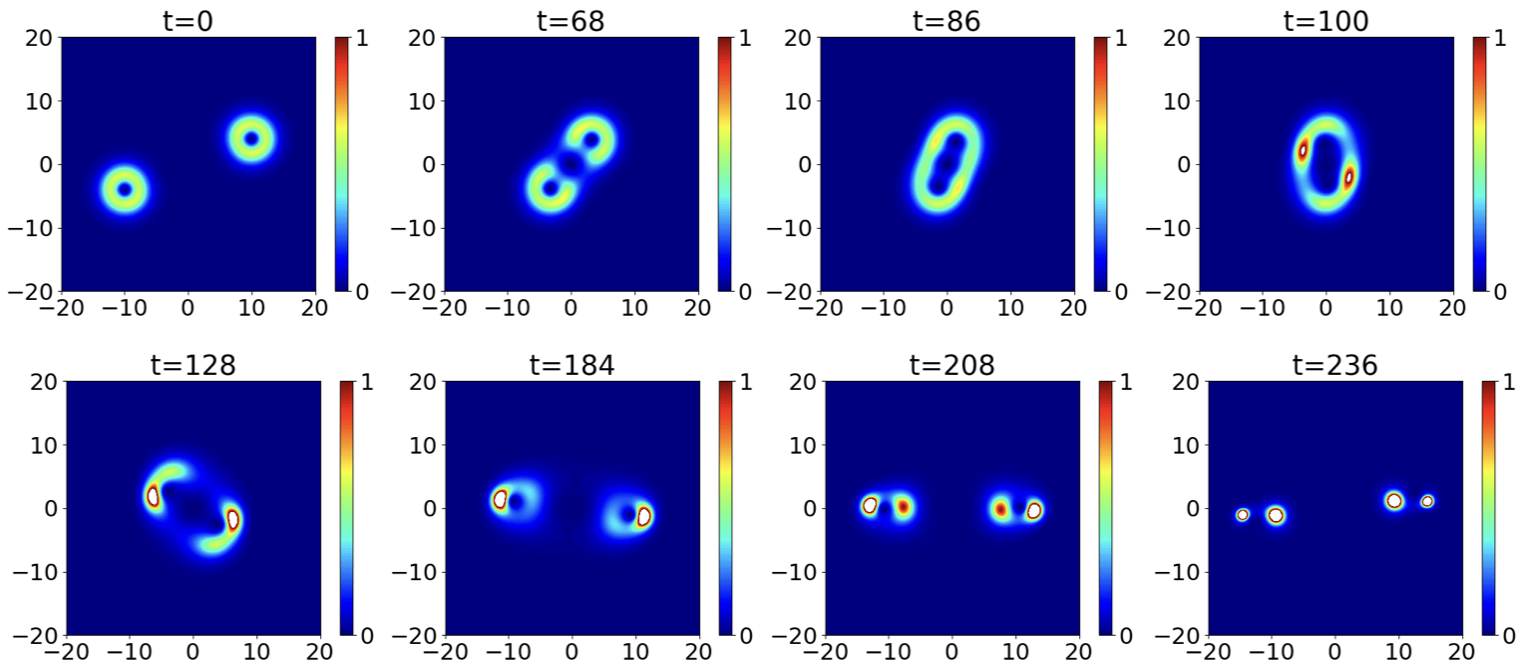}
    \caption{Energy density plots from field theory dynamics for the scattering of a pair of charge two Q-lumps with parameters $\lambda_1=\lambda_2=3, \ a=-10, \ b=-4, \ \chi=0, \ v=0.1.$}
               \label{4b4chi0}\end{center}\end{figure}

The initial condition (\ref{ic4}) with $\lambda_1=\lambda_2$ and $\chi=\pi$ satisfies $W(-z)=-W(z)$, and hence all the scatterings investigated so far in the charge four sector display a cyclic $C_2$ symmetry.
If $\lambda_1=\lambda_2$ and $\chi=0$, then again there is a $C_2$ symmetry, but this time realized as $W(-z)=W(z)$. The result of changing the phase from $\pi$ to zero can be illustrated by repeating the simulation presented in Fig.\ref{ex4b054}(c) and Fig.\ref{4b4}, but with the new phase. This produces the scattering displayed in Fig.\ref{sb22}(a) as a multiple exposure plot, and in Fig.\ref{4b4chi0} as energy density heat maps, with m06.mp4 the corresponding movie. Comparing Fig.\ref{4b4} and Fig.\ref{4b4chi0} reveals that the change in phase makes a considerable difference to the intermediate configuration that is formed, and subsequently to the outgoing Q-lumps. The deformation to each charge two Q-lump is now much stronger, with the amplitude of oscillation large enough that individual Q-lumps of different sizes are visible at some points in the oscillation, although the perturbation is not strong enough to yield fission, in which the Q-lumps would remain well-separated for all subsequent times.

The $C_2$ symmetry in the charge four scattering can be broken either by changing the initial relative phase so that $\chi\notin\{0,\pi\}$, or by choosing different initial sizes for the Q-lumps. An example of the first possibility is presented in Fig.\ref{sb22}(b), where the relative phase is taken to be $\chi=\pi/2$, and the second possibility is realized in Fig.\ref{sb22}(c), where $\lambda_1=2$ and $\lambda_2=4$, with $\chi=\pi$. These plots illustrate that, generically, a head-on collision results in fission that produces individual Q-lumps with a variety of speeds, scattering angles and sizes.

There is a general issue for Q-lumps, regarding the relevance of restricting dynamics to only finite energy solutions, because of the following reasoning. Any stationary Q-lump solution with infinite energy in the charge $N$ sector can be associated with a finite energy stationary Q-lump solution in the charge $N+1$ sector, via the addition of an extra Q-lump that can be placed arbitrarily far from any of the other existing Q-lumps. Explicitly, the process may be represented by the formula
\be
   W=\frac{\alpha_{N-1}z^{N-1} +\cdots +\alpha_1 z+\alpha_0}
   {z^N+\beta_{N-1}z^{N-1} +\cdots +\beta_1 z+\beta_0}e^{imt}-\frac{\alpha_{N-1}}{z-\mu}e^{imt},
   \label{ratmapadd}
   \ee
for a finite energy Q-lump with charge $N+1$. Here $\mu$ is a positive real parameter, that can be made arbitrarily large to move the extra Q-lump far from any of the existing Q-lumps that are contained within the infinite energy solution given by just the first term of (\ref{ratmapadd}).
This shows that for any infinite energy stationary Q-lump solution ${\cal S}$, and any choice of a compact region $\Omega$ of the plane, there is a finite energy stationary Q-lump solution $\widetilde{\cal S}$ that approximates ${\cal S}$ in $\Omega$ to any desired level of accuracy. In terms of local dynamics, it therefore seems difficult to argue that certain configurations must be ignored because they have infinite energy when considered over the whole plane.

\begin{figure}[!ht]\begin{center}
    \includegraphics[width=0.9\columnwidth]{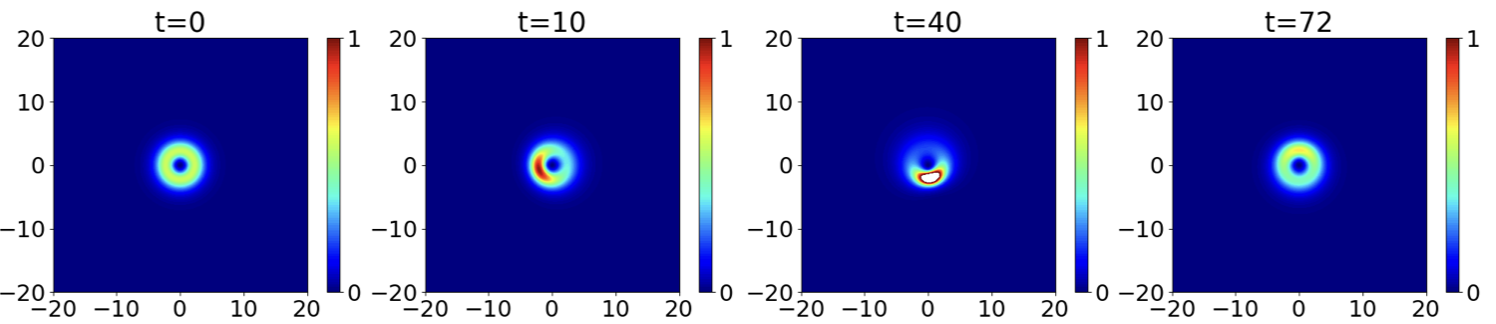}
    \caption{
Energy density plots from field theory dynamics for a perturbed charge two Q-lump with parameters $\lambda=3$ and $\epsilon=0.1$.}
    \label{2ax}\end{center}\end{figure}

A concrete example of the above general idea is provided by perturbations of the axially symmetric charge two Q-lump. Perturbed oscillating charge two Q-lumps appeared in some of the charge four scattering events presented above. These configurations cannot be studied within the moduli space approximation as isolated charge two Q-lumps, because the relevant stationary solutions have infinite energy. However, they can be studied using field theory simulations. Consider an initial condition taken from the field
\be
W=\frac{\lambda^2+z\epsilon t}{z^2}e^{imt},
\label{pert2}
\ee
where $\epsilon$ is a real parameter that induces a symmetry breaking perturbation of the axially symmetric charge two Q-lump, without splitting it into a pair of charge one Q-lumps. The resulting dynamics is shown as the energy density plots in Fig.\ref{2ax}, and as the movie m07.mp4, for the example with parameter values $\lambda=3$ and $\epsilon=0.1.$ It can be seen that the distorted configuration rotates and oscillates but does not separate into individual Q-lumps. Note the similarity between the deformed Q-lump at $t=40$ in Fig.\ref{2ax} and the pair of deformed Q-lumps in Fig.\ref{4b4chi0} at $t=184$. This confirms that field theory simulations are useful in studying Q-lump dynamics, even when the moduli space approximation is not applicable because the field configuration has infinite energy when extended to the full plane. 

Once the constraint of finite energy is removed, there are more possibilities for Q-lump scattering, particularly in the charge two sector. The relative phase can be unfrozen from $\chi=\pi$, and the pair of Q-lumps can be given different initial sizes. The appropriate initial condition is taken from the field
\be
W=\bigg(\frac{\lambda_1}{z-a-vt-ib}+\frac{e^{i\chi}\lambda_2}{z+a+vt+ib}\bigg)e^{imt}.
\label{icinf}
\ee
\begin{figure}[ht]\begin{center}
    \includegraphics[width=0.8\columnwidth]{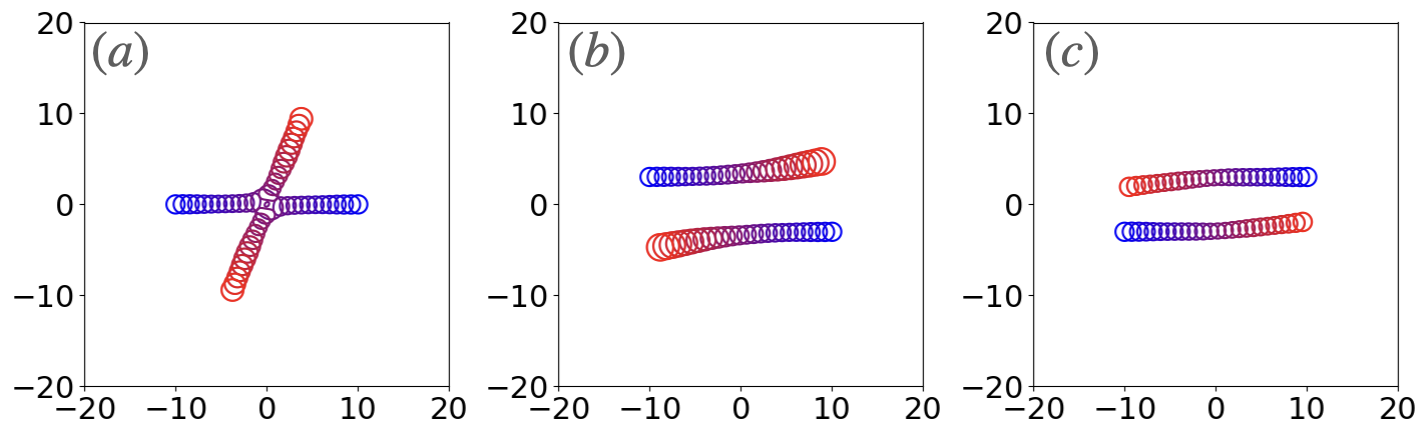}
    \caption{
Field theory simulations of the scattering of a pair of charge one Q-lumps with $\lambda_1=\lambda_2=1, \ a=-10, \ v=0.1, \ \chi=0$ and 
      (a) $b=0$;
      (b) $b=3$;
(c) $b=-3$.
    }\label{exch2chi0}\end{center}\end{figure}

The field theory simulations presented in Fig.\ref{exch2chi0} illustrate how the previous scattering with $\chi=\pi$, for example as shown in Fig.\ref{mex2}, is modified by changing the phase to $\chi=0.$ The other parameters are taken to be $\lambda_1=\lambda_2=1, \, a=-10, \, v=0.1$, with three different values for the impact parameter $b$. The head-on collision ($b=0$) in Fig.\ref{exch2chi0}(a)
shows a scattering angle less than $\pi/2$, in contrast to the $\chi=\pi$ scattering, where the scattering angle always lies in the interval $(\pi/2,\pi).$ As the initial speed decreases, the scattering angle increases, and is close to $\pi/2$ for $v=0.05$, for example. A positive impact parameter, Fig.\ref{exch2chi0}(b) with $b=3$, shows repulsion, and a negative impact parameter Fig.\ref{exch2chi0}(c) with $b=-3$, reveals attraction. This agrees with the previous $\chi=\pi$ situation, although these forces now have less influence as the interaction is minimal. Otherwise, as expected from the arguments given above, there is little within these results to indicate a significant difference between finite and infinite energy dynamics at the local level.

\begin{figure}[ht]\begin{center}
    \includegraphics[width=0.8\columnwidth]{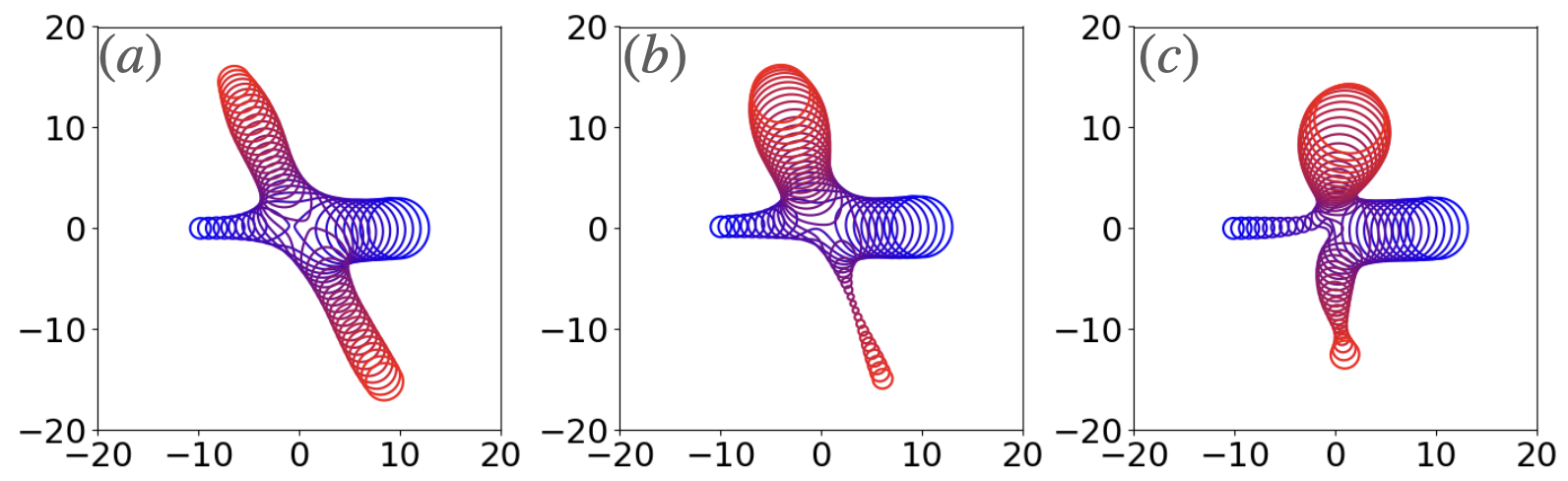}
    \caption{
Field theory simulations of the scattering of a pair of charge one Q-lumps with 
  $\lambda_1=1,\ \lambda_2=3, \ a=-10, \ b=0, \ v=0.2$ and
      (a) $\chi=\pi$;
      (b) $\chi=\pi/2$;
      (c) $\chi=0$.
          }
          \label{ex2l13}\end{center}\end{figure}

Finally,  Fig.\ref{ex2l13} presents some examples of the head-on scattering of unit charge Q-lumps, where the pair of Q-lumps have different initial sizes. The parameter values for these simulations are  $\lambda_1=1,\, \lambda_2=3, \, a=-10, \ b=0, \, v=0.2$, with three different values of the initial relative phase $\chi=\pi,\pi/2,0.$ This provides a clear illustration of the significance of the initial phase in determining the outcome of the scattering, with dramatic changes in the scattering angles, sizes, and amplitudes of size oscillations, as $\chi$ is varied.
        
 \section{Conclusion}\quad
 The moduli space approximation has been used to study Q-lump scattering, extending previous studies in the charge two sector to higher charges by imposing cyclic symmetries that restrict the motion to a 4-dimensional manifold. Field theory simulations of Q-lump scattering have been performed for the first time, with results that show an excellent agreement with moduli space dynamics. A range of exotic scattering events have been presented that include Q-lump fission and double scattering phenomena. Field theory simulations have also been applied to situations where the moduli space approximation is not applicable, revealing that considerations of finite energy are not particularly relevant in the study of local Q-lump dynamics.

There are several directions in which this work could be extended. For example, the spatial plane could be replaced by a compact manifold, such as a torus, as a natural way to unfreeze the moduli that are fixed in the planar case by finite energy considerations. Q-lumps can also be generalized to systems in which the target space is a K\"ahler manifold with a continuous isometry that has at least one fixed point \cite{Ab}. It might be interesting to investigate Q-lump dynamics in such systems, to see if any new features appear. 

Dyonic instantons provide gauge theory analogues of Q-lumps in (4+1)-dimensions, where moduli space dynamics has been applied to study scattering in the charge two sector \cite{PZ,AS}. Higher charge investigations could be performed by imposing symmetries, as in the present paper for Q-lumps, and perhaps field theory simulations are just about in reach with current computing capabilities.

Finally, the study of spinning topological solitons in modified $\sigma$-models is of general interest. In particular, spinning Skyrmions in (3+1)-dimensions can be regarded as approximations to nucleons with quantized spin, so the classical scattering of spinning Skyrmions is relevant to the study of nucleon-nucleon scattering \cite{Sc,GP,Ma2,FM}. Q-lumps provide a simple lower-dimensional analogue of this situation, so a detailed understanding of their dynamics may provide some insight into the more complicated Skyrmion system.

\end{document}